# The nanomorphology of cell surfaces of adhered osteoblasts

C. Voelkner[1,2], M. Wendt[1,2], R. Lange[1,2], M. Ulbrich[1,2], M. Gruening[1,3], S. Staehlke[1,3], J.B. Nebe[1,3], I. Barke[1,2] and S. Speller[*,1,2]

Address: [1]Department Science and Technology of Life, Light and Matter, University of Rostock, Albert-Einstein-Str. 25, 18059 Rostock, Germany. [2]Institute of Physics, University of Rostock, Albert-Einstein-Str. 23, 18059 Rostock, Germany. [3]Department of Cell Biology, Rostock University Medical Center, Schillingallee 69, 18057 Rostock, Germany.

Email: Sylvia Speller – sylvia.speller@uni-rostock.de
* Corresponding author

## Abstract

Functionality of living cells is inherently linked to subunits with dimensions on the nanoscale. In case of osteoblasts the cell surface plays a particularly important role for adhesion and spreading which are crucial properties with regard to bone implants. Here we present a comprehensive characterization of the 3D nanomorphology of living as well as fixed osteoblastic cells using scanning ion conductance microscopy (SICM) which is a nanoprobing method largely avoiding forces. Dynamic ruffles are observed, manifesting themselves in characteristic membrane protrusions. They contribute to the overall surface corrugation which we systematically study by introducing the relative 3D excess area as a function of projected adhesion area. A clear anticorrelation is found upon analysis of ~40 different cells on glass as well as on amine covered surfaces. At the rim of lamellipodia characteristic edge heights between 100 nm and ~300 nm are observed. Power spectral densities of membrane fluctuations show frequency-dependent decay exponents in excess of -2 on living osteoblasts. We discuss the capability of apical membrane features and fluctuation dynamics in aiding assessment of adhesion and migration properties on a single-cell basis.

# Introduction

Osteoblasts are bone-mineralizing cells situated inside the matrix boundaries of the osteoid. They adhere to and spread on a wide spectrum of pristine and coated material surfaces such as titanium and polyallylamine [1–3], gelatin-nanogold [4], polyelectrolyte and RGD peptides [5] or extracellular matrix proteins [3, 6, 7]. Especially in the unphysiological situation of material surfaces for permanent implants, settling and the swift formation of large adhesion interface areas are desired. For that reason a variety of surface coatings has been assessed, relevant parameters may be the zeta-potential and pre-adsorbed cell adhesion protein from the serum of the medium [2, 8]. Before the osteoblasts start their adhesion and spreading program they settle on the material surface. Apart from cell-biologic parameters the adhesion interface area and the speed of its formation are insightful as a measure of surface bio-compatibility with regard to osteoblastic cells [9].

Dynamic remodeling of the cytoskeleton is the basis for shape adaptation and migration for many mammalian cell types [10, 11]. Migrating and spreading cells form flat, actin-supported, organelle-free regions, referred to as lamellipodia, and other features which may expand their attachment area [12]. A physical coupling of adhesion molecules to the actin polymerization machinery has been determined [13]. In the course of adhesion-related cell processes considerable rearrangements take place on the nanoscale in and on cells. Some of them are difficult to address by optical imaging methods due to limited resolution or unduly high light exposure. Scanning Probe Microscopy is an option to study the membrane surface nanoscopically, without dye labeling or laser light exposure.

In Scanning Probe Microscopy a nanoprobe is kept at a constant distance from the sample surface by maintaining a local interaction signal constant via a feedback loop [14]. If the interaction signal is a force, pressure is applied to the sample. This is the case with atomic force microscopy (AFM) giving rise to substantially depressed apparent heights on living and fixed cells [15]. Typically, mammalian cells exhibit Young´s moduli in the regime of 1 to 10 kPa while AFM probe pressures may correspond to 10 MPa, e.g. assuming 1 nN loading force and 5 nm tip radius. Though the resulting artificial membrane depression depths on cells amount to about half a micrometer already at 1 nN, AFM is useful to measure the cortical actin network underneath the membrane [16].

Regarding the investigation of the membrane itself, an intricate issue of applied forces is that cellular responses may get triggered and thus the native state of the membrane might be concealed. A localized ion current flowing through a nanopipette probe represents a suitable non-invasive interaction which is exploited in Scanning ion conductance microscopy (SICM) [17–19]. Therefore, it is well suited to probe soft and responsive surfaces such as those on living cells. The applied pressure is only a few hundred Pa and results from hydrostatic pressure of the fill level of the nanopipette [20]. The ion current drops with probe-sample approach, because the effective area for the ion trajectories becomes smaller, referred to as current squeezing. SICM is the only method capable of nanoscopic three-dimensional spatial resolution on living cells, without application of dye labels or other modifications.

Though SICM was developed early [17], initially it was not much exploited until the further deployment of the method for a number of mural and human cell line types [21]. Meanwhile living cell nanomorphologies have been acquired for a number of mammalian cell types such as cardiomyocytes, fibroblastic cells, neurons, renal and epithelial cells [22–25]. Osteoblast surfaces have not been addressed, neither fixed nor live. We investigate the nanomorphologies of osteoblast-like cells (MG-63) adhered on glass and amine functionalized surfaces in living and fixed state. Our results include characteristic sheet-like protrusions, so called ruffles. Their appearance on the osteoblast cell rims mostly vanishes when a large adhesion area is established, resulting in a smooth apical plasma membrane surface. Several other morphological and dynamic parameters are evaluated e.g. cell edge heights, membrane surface roughness and membrane fluctuations and discussed with respect to cellular functions.

## Results and Discussion

In Figure 1 we show a typical overview SICM image of a whole osteoblast in the fixed state. The morphology is polar, i.e. a flat region (lamellipodium) at the "leading" side and a bulkier region containing the nucleus at the trailing side have been formed. The height of the bulky side of the shown cell is 8 µm while that of the lamellipodium is 500 nm. However, the dimensions and the extent of polarity vary among the cells (not shown). With time the adhesion interface increases and polarity gets less distinct. Furthermore, the cell exhibits a pronounced elevation of apparently 2 µm and with a width of 1 µm (see blue graph in Figure 1b). Such a SICM signature is compatible with that of a primary cilium at 1% ion current reduction [26].

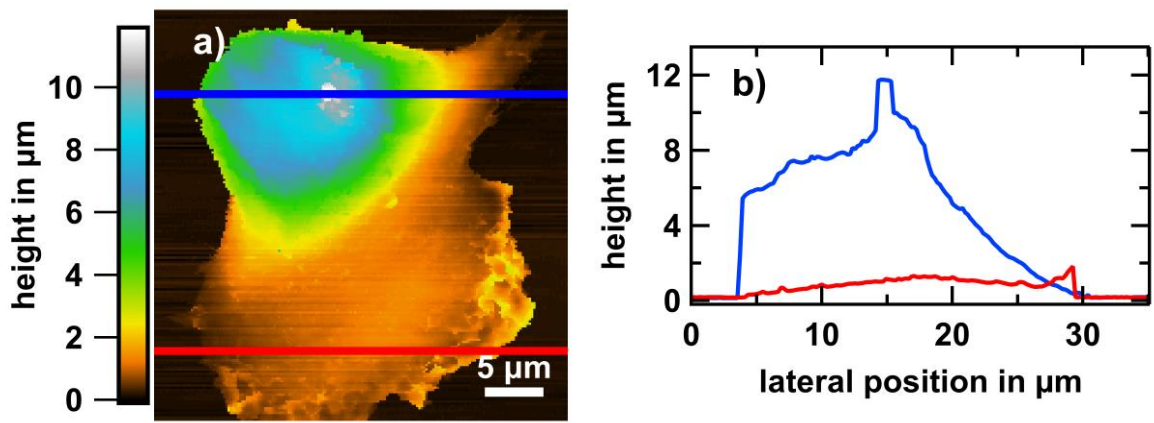

Figure 1: SICM analysis of one fixed MG-63 osteoblastic cell. a) Overview morphology of the cell on glass after 24 h culturing. The quite polar morphology shows that the cell was migrating towards the bottom of the image. b) Exemplary height profiles of the trailing edge (blue) and the leading edge (red) along horizontal lines indicated in a).

**Dorsal Ruffles**

The morphologies of osteoblast membranes reveal leaf-like protrusions, so called ruffles. They generally cover the whole cell surface. They are quite thin (similar to the thickness of filopodia ~100 nm) and flexible. Figure 2a shows an example of a typical SICM topography from the border region in the live state (see Figure 2b for the corresponding optical microscopy image). Features with lateral dimensions of approximately 1 µm x 0.8 µm protruding 100 – 300 nm from the surrounding (see Figure 2c) at a density of 0.3 – 0.5 features per µm$^2$ are observed. The characteristic time scale of the dynamics of the leaflets is small compared to the acquisition time leading to temporal undersampling. This results in distortions, particularly evident in Figure 2e where the ruffles appear as blurred, bright spots.

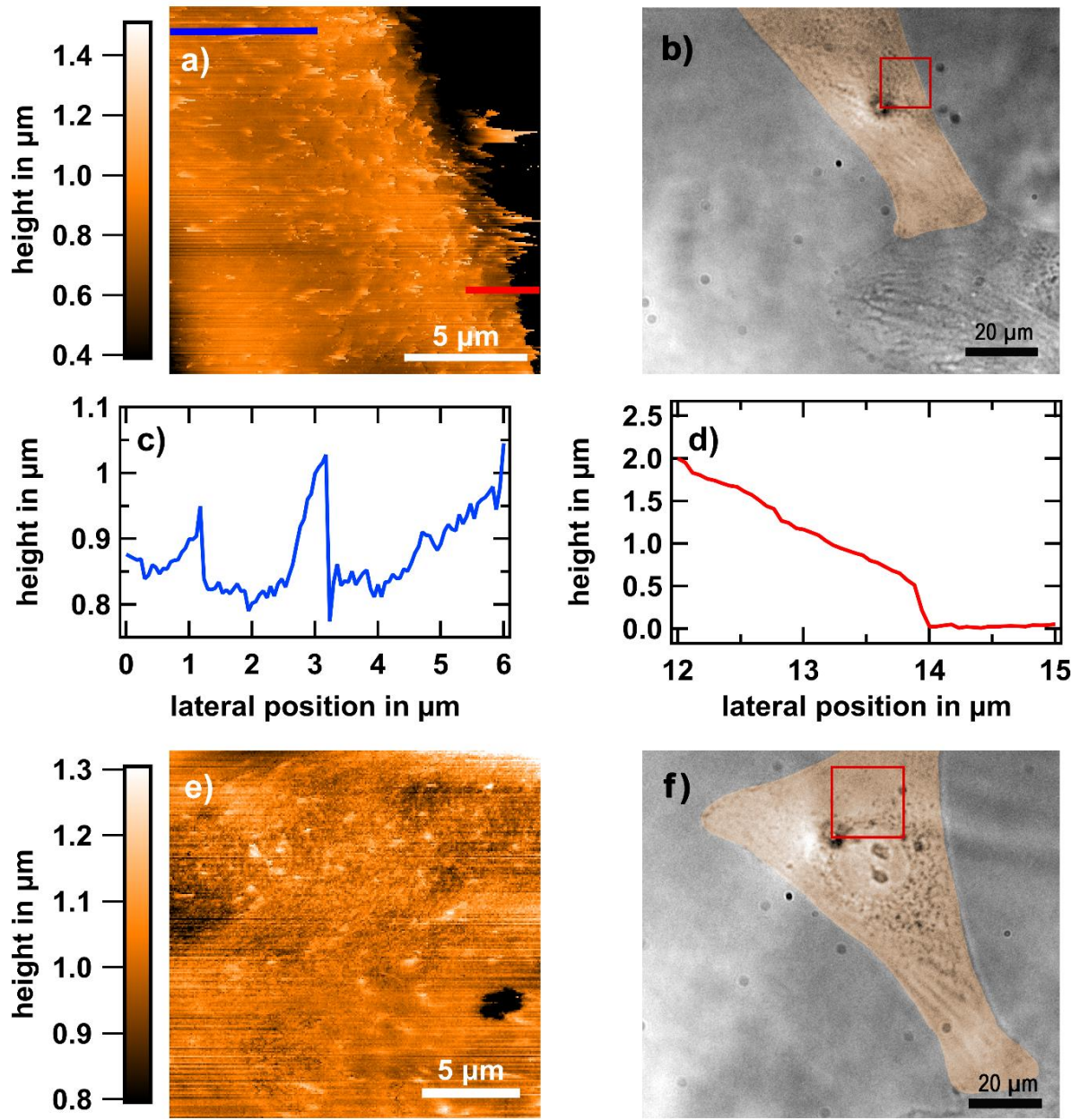

Figure 2: Living osteoblast adhered to glass after 24 h. a), e) SICM-topography images. The acquisition time was 68 minutes per image. Cell membrane protrusions, so called ruffles can be seen as bright spots. b), f) Corresponding optical images of the living cell taken prior to the SICM measurements. The cell is digitally stained in orange for better visibility. The red marked areas were measured with the SICM. a) Sheet-like protrusions (ruffles) are visible that mostly orient to the right side. c) Height profile along blue line indicated in a). Typical feature heights are several hundred nanometers.
d) Line profile at the cell rim, as indicated in a). A sudden jump from cell surface to the flat glass surface is visible, which we refer to as the step edge height (here roughly 500 nm).

This becomes obvious if the live-cell dynamics is suppressed upon fixation of the osteoblasts with 4% paraformaldehyde (PFA). Figure 3a shows an example of a respective SICM topography. Now the ruffles exhibit a clearer shape and resemble similar features observed by electron microscopy [27, 28]. Our data reveal that the ruffles are ragged, they resemble fins, and they confine an acute angle towards the membrane. Frequently, they show twisting. In the fixed state their angle is not very variable, they do not show flapping if the fast scan direction is reversed.

Similar sheet-like structures have been observed for dendritic cells. It has been shown that such veils function as catchers for T-cells [29]. Ruffles have been described on osteoblasts upon exposure to parathyroid extract [30], and after internalization of polymer or metal particles [31], fibroblastoid cells [27, 32], breast cancer cells [28], and on keratinocytes ruffles. In general these features are associated with migration, receptor internalization, and macropinocytosis [28, 33, 34]. Membrane ruffling is regulated by a distinct signaling pathway [35] and the supporting actin is denser and more cross-linked [36] compared to flat membrane regions. Three types of ruffles are discriminated: (linear) dorsal, peripheral and circular dorsal ruffles [33]. Colocalization with hyaluronan synthase has been found using a breast cancer cell line (MCF-7) [28]. The ion current error channel images of living versus fixed cells provide insight into the protein content in the ruffle volume. Usually the error current in SICM is higher on hard materials than on soft (see e.g. Figure 3b, white area on the right side, corresponding to glass). The reason is that the current versus distance curve is steeper on hard than on soft materials [37]. The fixed ruffles exhibit an extremely low error (Figure 3b), lower than the surrounding membrane. PFA usually increases Young's modulus of cell surfaces, for instance from 3.5 kPa to 18 kPa on fibroblasts [37]. Therefore, the lower ion current error appears counterintuitive from point of view of material properties. However, considering the fin or springboard morphology of ruffles, the elasticity may not result from pure material property changes but also from the flexible shape. Thus, the extremely low error probably points towards lower stiffness of the ruffle structure, even at harder (fixed) protein content. Note that for living cells such contrast is absent as can be seen in Figure 3c and d.

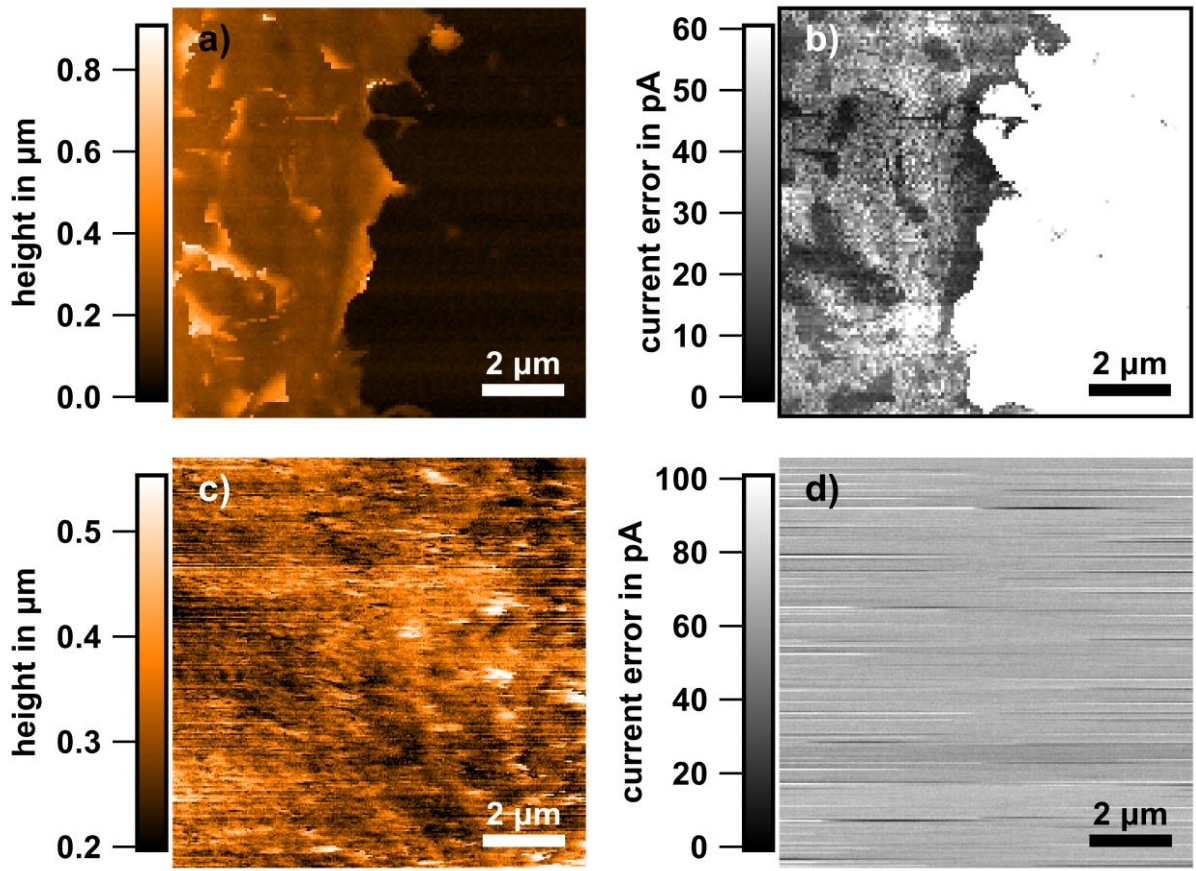

Figure 3: Comparison of living and fixed MG-63 cells by SICM. a) Osteoblast adhered (3 h) to glass - after fixation. SICM topography image at the edge of a cell. The ruffles are clearly visible. b) The corresponding current error image. Note that the ruffles appear to lead to smaller error-values resulting in a contrast in the error map. c) SICM topography of a living osteoblast adhered to glass (24 h). d) The corresponding current error map shows that there is no error contrast in the live state.

Figure 4a shows an example of a peripheral ruffle. Peripheral ruffles sometimes are attributed to loosened lamellipodia. Such loosening and its subsequent retraction towards the cell body has been shown for keratocytes [36]. Either the attachment of the lamellipodium via integrin was not successful or such processes may serve the cell to attain flexibility of structures and thereby responsiveness. Dynamic remodeling of structures which requires a continuous balance between assembly and disassembly, is well known for cytoskeletal fibers.

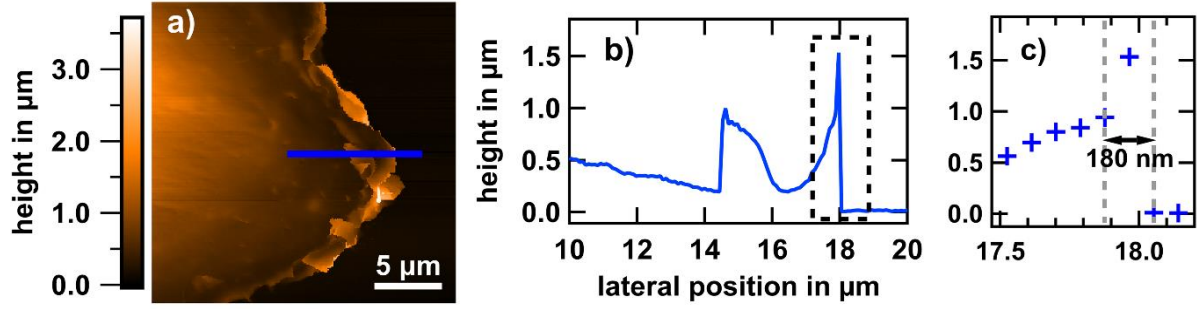

Figure 4: SICM analysis of a fixed cell adhered for 24 h to a 10 nm Au layer. a) SICM topography of cell's peripheral ruffles. Such ruffles can exhibit extensions of more than one μm as shown by the blue marked location. b) The corresponding line profile. c) The dashed section of the plot separately shown. Note that the upper thickness limit of this peripheral ruffle is 180 nm, assuming that it forms an angle of 90° with respect to the glass surface.

To quantify the cells' excess membrane, we fixed the cells already 3 h after adhesion. The reason is that most of the spreading happens in the early phase after initial adhesion [38]. In order to quantify the extra membrane, associated with dorsal ruffles, we define the excess surface $A_{exc}$ as the difference between the effective and the projected surface area ($A_{eff} - A_{proj}$). The effective surface is the undulated surface area of the three-dimensional function z(x,y) as determined from SICM topographies

$$A_{eff} = \iint \left[ \left( \frac{\partial z(x,y)}{\partial x} \right)^2 + \left( \frac{\partial z(x,y)}{\partial y} \right)^2 + 1 \right] dxdy$$

and the projected area is the frame or base area $A_{proj} = \iint dxdy$. (see Figure 5b). The relative excess surface is then $A_{rel,exc} = \frac{A_{eff} - A_{proj}}{A_{proj}}$. Figure 5a shows the relative excess surface as it correlates with the adhesion area of the whole cell $A_{rel,exc}(A_{adh})$ [39]. The total adhesion area $A_{adh}$ is determined via optical microscopy images. Relative excess surfaces between 5% and 60% are found assessing ~40 cells. Although the data is substantially scattered, we can observe a clear anticorrelation, meaning $A_{rel,exc} \sim A_{adh}^{\alpha}$ with $\alpha = -1.13 \pm 0.25$ which was obtained by least square fit and indicated by the red curve in Figure 5a. This means larger adhesion interface area comes along with a smaller excess surface, i.e. a lower number of or less extended ruffles. This could mean that excess membrane has been "consumed" in more anisotropic or spread cell shapes to compensate for the increased surface-to-volume ratio in the course of spreading. Ruffles could serve as a membrane reserve prior to formation of large adhesion interfaces. This would be compatible with the observation that there are less ruffles found on lamellipodia or rim regions versus more central regions on the cell. For keratocytes an anticorrelation between ruffling and lamellipodia persistence has been reported [36], supporting this scenario. Another option is the ruffles, excess membrane, being a sign of former unsuccessful spreading or deliberate lamellipodia loosening, e.g. for the purpose resumption of migration [13].

Having a look at the substrate-specific data, cells with extremely large adhesion interface areas (>2600 µm²) are only found on PPAAm layers. Nevertheless, the span of measured adhesion areas is large. This may indicate that the PPAAm layers on our glass surfaces are heterogeneous, or that cells do not always respond to it. Another origin of the large spreading may be, that depending on their stage in the cell cycle the adhesion program may get more or less priority. It has been shown that isolated cells do not synchronize their cell cycle in contrast to cells in groups [40] or tissue [41, 42], i.e. each cell may be in a different stage. This may contribute to different spreading speeds as observed on PPAAm. According to Pu et al. [43] MG-63 osteoblast-like cells are at 62%, 18.6%, and 19,4% in G1,S,and G2 phases, respectively. In average (analyzing 44 cells and taking 3 samplings on each cell) the excess surface on glass (0.25 ± 0.03) is larger than on PPAAm (0.17 ± 0.04). A slightly positive zeta-potential, such as measured for PPAAm on titanium, has been figured out being beneficial for the spreading speed [2, 44]; our tentative observation of larger maximal adhesion areas on PPAAm than on glass substrates is compatible with these earlier findings. Our result is not compatible with [45], where ruffles have been predominantly observed on primary rat osteoblasts cultured on bioglass, which is less negative in zeta-potential, compared to quartz glass.

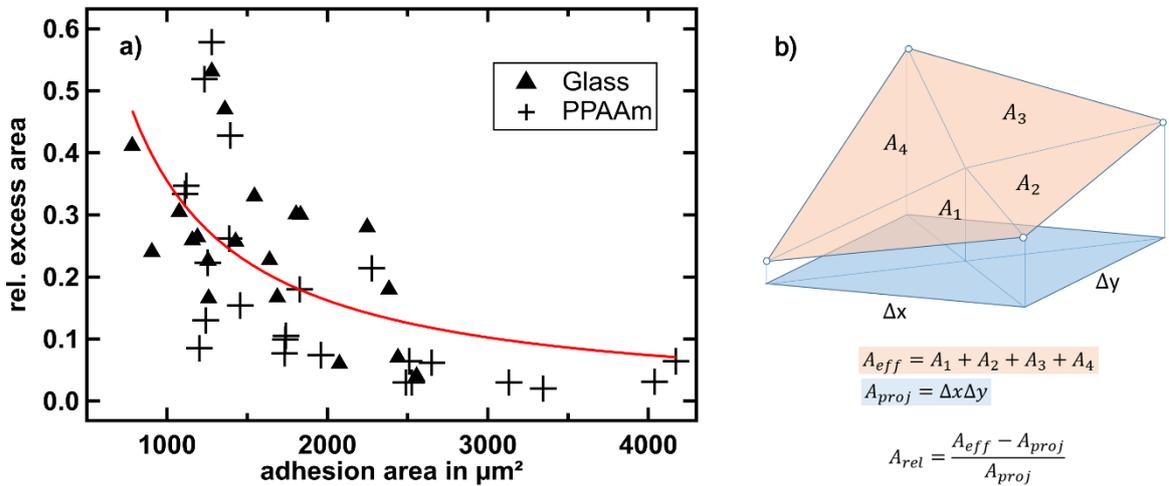

Figure 5: a) Correlation plot of the relative excess surface (ordinate) and the total adhesion interface area (abscissa) 3 h after seeding. A total of 4 different cell passages has been investigated. Triangles mark values for cells on glass and crosses those for cells on the PPAAm coating of glass. On each cell 3 edge locations were chosen randomly. The values depicted in the graph are the arithmetic mean of the corresponding relative excess areas. An anticorrelation becomes evident (red graph). b) Scheme of the effective surface area (orange) and the corresponding projected surface area (blue). The white dots represent height -data points.

## Membrane holes and smaller protrusions

Though the ruffles are the most prominent feature we observe on the osteoblastic cells, further membrane features such as holes and circular protrusions are encountered. Circular tail-like protrusions have been sometimes observed along with the ruffles (see Figure 6a). They measure below 100 nm in length. It is difficult to conclude whether those are microvilli or filopodia, or just an early form of ruffles. Microvilli, among other, serve to enhance the exchange of substances with the extracellular medium (absorption, secretion) while filopodia are used to explore the environment of an adhering cell on a surface, especially if it exhibits some roughness or edges.

Occasionally we observe depression or hole features in the membrane. They are a couple of 100 nm up to one µm deep and mostly occur in close vicinity of extensive ruffles (Figure 2e and Figure 6b). These features may relate to (macro)pinocytosis, i.e. the cells take up smaller or bigger volumes of the extracellular medium. Indeed ruffling, especially its circular form has been associated to pinocytosis [28, 33, 34]. Linear ruffles may be regarded as precursory structures to circular ruffles, and latter turn in to perform pinocytosis [33, 34]. The lower incidence and the different dimensions of the circular ruffle type (observed as hole with more or less pronounced "roll collar") is compatible with the assumption that they are much more transient, i.e. they occur transiently before the extracellular medium is taken up and close shortly after. In SICM observations we found 3 times more (on ~50 cells) structures resembling circular dorsal ruffles which is consistent with our SEM observations, where we found only one circular dorsal ruffle among 20 cells (not shown).

Rarely, pancake-shaped structures are seen on the membrane such as shown in Figure 6c. Since they also measure ~200 nm in height these elevations are definitely no lipid rafts. They could be ruffles which are not sideways anchored to the membrane but somewhere more central underneath; note that the nanopipette of the SICM has no access to hollows underneath other surfaces.

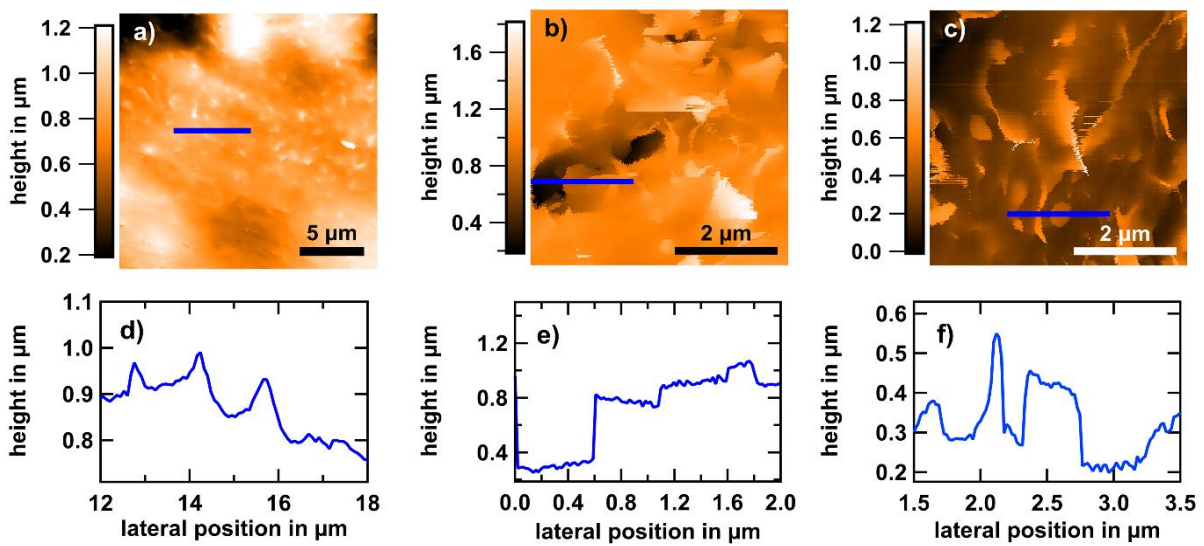

Figure 6: Other membrane features observed by SICM on fixed osteoblastic cells on glass (24 h). a) Circular tail-like protrusions in the vicinity of the cell nucleus. These tail-like features are characterized by heights of less than 100 nm and seem to show no flapping which results in almost orthogonal orientation with respect to the cell membrane. b) Depressions or holes in the cell membrane of ~500 nm depths. Similar features in literature are attributed to macropinocytosis events [46]. They may result from the collapse of circular dorsal ruffles. c) Pancake-like structures that were rarely observed shows plateau-like characteristics. The heights observed are similar to those of dorsal ruffles. d-f) The corresponding line profiles.

Lengthy, shallow wrinkles with periodicities around 500 nm occur on elongated cells or larger extensions of them (see Figure 7a). Since they come along with anisotropic shapes and the wrinkles then usually are aligned parallel to the long side of the cell or its extension, we suspect that they are related to the cytoskeleton underneath, i.e. the membrane may cling to the proximate fiber network.

Largely featureless regions, free of ruffles and other membrane structures, are scarce; they show waviness on the mesoscopic scale and 2D-rms roughness of 17 nm on the nanoscopic scale, as illustrated in Figure 7b.

We find it noteworthy that hardly any filopodia are expressed at cellular rims, while this type of extensions has been frequently observed on rough or microstructured surfaces.

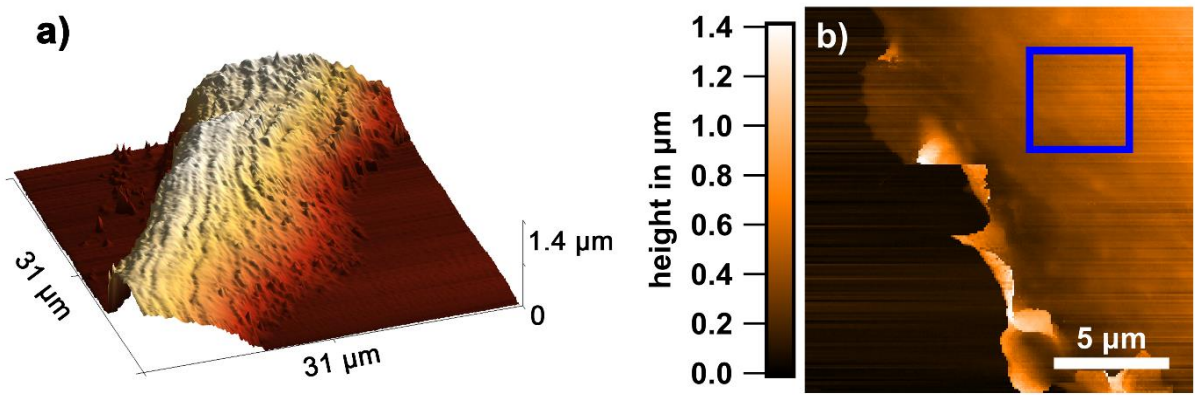

Figure 7: Further membrane features and apparently featureless regions observed by SICM. a) Pseudo-3D-view of a living osteoblastic cell on glass for ~5h. Lengthwise wrinkles that follow the macroscopic shape of the cell are developed. They protrude 100-500 nm from the cell surface. b) Topography image of a largely featureless membrane region on the edge of a fixed cell. The blue marked area was exemplarily chosen to determine the 2D-rms roughness value of about 17 nm. Note that a plane was subtracted from the dataset in advance to eliminate the tilt.

## Ruffles observed by SEM

Osteoblastic cells have been studied extensively by scanning electron microscopy (SEM) [31, 45, 30]. Electron microscopy can yield nanoscopic resolution, however requires invasive preparation such as critical point drying and deposition of thin Au layers. This enables the observation of cell surfaces containing membrane protrusion features; however, apart from [30] no such clear ruffles as observed by SICM are recognizable. If no Au coating was applied on the cells of the same osteoblastic cell line (MG-63) which were fixed after 3 h of adhesion on titanium and afterwards critical point dried, the ruffles stay clearer (see Figure 8). There are two issues towards this contrast discrepancy which cannot fully be excluded to play a role. By means of the critical point drying preparation method volume changes of the cell throughout the preparation procedure for SEM are believed to be largely avoided. This might nanoscopically not fully hold and some membrane protrusion features could get altered during cell death and drying procedure. Thus, an option to explain the difference between SEM and SICM images is that upon critical point drying the ruffles might be too fragile and partly be retracted or collapse. However, especially the attractive van der Waals forces might lead to attachment of ruffles on the membrane surface and hence their deformation in vacuum conditions. In contrast to that the attractive surface forces are simply screened in physiological medium by ions, leading to undistorted ruffles. Indeed, plenty of "wormy" structures are typically observed on osteoblastic cells in SEM images which could be bent microvilli or residual ruffles.

In summary, plasma membranes in the rim region tend to appear slightly smoother (less ruffled) on cells which have formed large adhesion interface areas. This may indicate that membrane supply is an issue at faster spreading.

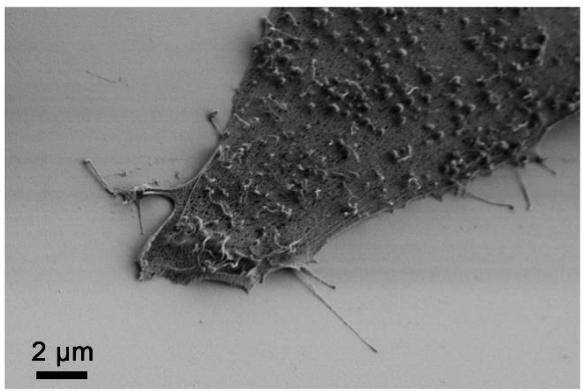

Figure 8: Scanning electron microscopy of an MG-63 osteoblast after 3 h adhesion to titanium. The ruffles on the cell membrane are visible. (FE-SEM SUPRA25, 1 kV, 30° angle, 100 nm Ti on Si wafer, fixation by 2,5 % glutardialdehyde (GA), acetone series, critical point drying).

# Rim edge heights

In Figure 9a we show a typical example of an osteoblastic cell's rim step edge height. Other MG-63 cells exhibit similar edge heights although the weighting of the different height regimes measured varies with the overall shape of the cell. The step edge heights have been systematically measured at the rim of the lamellipodia as well as at ordinary edges of the cells. Note, that the step edge height not only incorporates the cleft between cell and material surface but additionally the local height of the cell itself (see Figure 9b). Both parameters may vary independently, and from the three-dimensional cell morphology we can only determine the sum of cleft and cell height. At the higher edges (beyond one µm) the step edge slope is greater than 89° and there is hardly any curvature detectable on the cell before its height jumps to the substrate level. This could mean that the cell has a pronouncedly undermined configuration, compatible with the cleft distance rising at the rim of cells (see discussion below). The flatter cell edges do show slight curvatures on top and at the ledge side and slope angles measure down to 45°. The smallest heights at homogeneous regions measured about 100 nm (Figure 9a). Having in mind that the extracellular matrix (ECM) will take already about 60 nm [47] the curvature radius of the membrane would correspond to 20 nm. Values in this regime are in line with protein-aided cellular membrane curvatures. E.g. intracellular membrane tubes exhibit diameters of a few 10 nm, a maximum curvature is assumed to be generated via Bin/ampiphysin/Rvs (BAR) protein and corresponds to < 1Å [48, 49]. Sometimes we find rugged traces with only 60 nm height. Here we suspect that those may originate from the extracellular matrix's protein secretion left behind, material formerly present within the now resolved cleft. Either this protein becomes exposed because the lamellipodium has folded away from the surface (to become a peripheral ruffle) or part of the cell has retracted from that region in the course of migration. The smallest heights beyond these minimal heights measure around 100 nm, which we attribute to lamellipodia heights including the cleft. Apart from that, we find quite high step edges of ~ 1 µm and beyond which we attribute to ordinary or trailing edges of osteoblastic cells (Figure 9a).

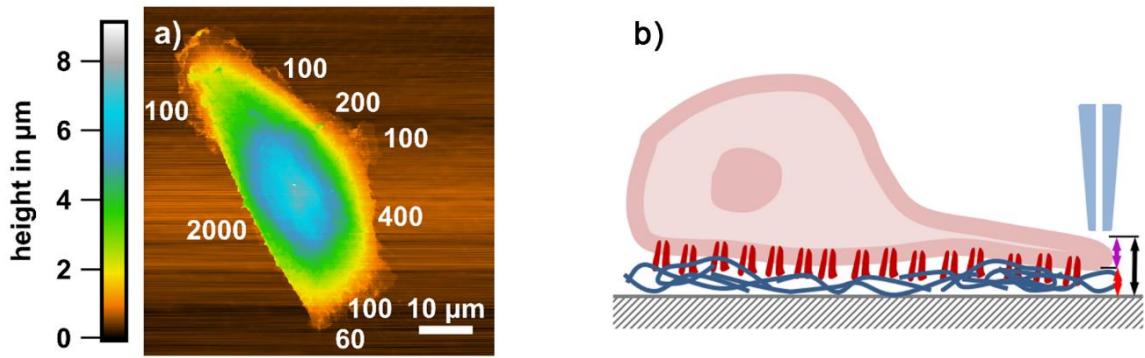

Figure 9: Edge height analysis. a) Topography of a fixed osteoblast on a 10 nm Au layer on glass with indicated results (in nm) of step edge heights analyses at the rim region. Cells show vastly differing edge heights, depending on the macroscopic location. However, the transition from regions of big edge heights to smaller ones is rather smooth. b) Scheme of the measuring principle of the step edge heights. We determine the overall height of the edges (black arrow), which contains the cell thickness at the rim (purple arrow) as well as the cleft between substrate and basal cell membrane (red arrow).

An elegant method to measure the cell – surface cleft at confocal resolution is metal-induced energy transfer (MIET) where the fluorescence lifetime map of a membrane label is transferred into a cleft distance map [47]. The cell is adhered to a transparent metal layer and cleft distances of 67 nm +/-7 nm have been determined for fibroblast-like cells (A549). Interestingly, the cleft spatially varies by more than 40 nm along the adhesion interface and by more than 70 nm in space and time. At the peripheral region higher cleft distances have been observed. These values do include only the extracellular matrix, because the dye is situated in the plasma membrane. Our step edge heights are compatible with these results and point towards lamellipodia thicknesses in the regime of 100 to 200 nm for osteoblastic cells.

An alternative approach to measure the thickness of lamellipodia is to measure the thickness (or better the width) of "upright" peripheral ruffles, i.e. former lamellipodia, such as shown in Figure 4a. Note that the aspect ratio of the nanopipette probe is rather high and steep slopes to almost 90° can be reproduced. Such analyses yield thickness values of 180 nm. This can be seen as an upper limit for the height of a fully expressed lamellipodium. Thus, the heights at the osteoblastic cell's rim edges vary between 100 nm and more than 2 µm along their circumference, and they are similar on both substrate types.

## Membrane fluctuations

Since topographic measurements were shown not to be sufficient to evaluate dynamic processes like migration (see Figures 2a,e), we focused on inspecting local apical membrane fluctuations. They might contain valuable information about their origin and give new insights in intracellular dynamics at the leading edge of cells and thus enable assessment of e.g. cellular migration properties of the substrate. Especially, forward-backward oscillations, "actin waves" being important for lamellipodia formation have already been shown to dominate membrane fluctuations with characteristic timescales of ~10 seconds and amplitudes of more than hundred nanometer [50, 51]. To investigate the apical fluctuations in view of cellular migration, we selected a few positions on the cell (somewhere between nucleus and cell edge) and acquired time traces either of current or of height variations. The acquired signal covers a frequency range similar to that of the pre-amplifier's bandwidth of up to 1 kHz. For the extraction of root mean square (rms) fluctuation amplitudes the statistical standard deviation of the SICM height values (~$10^4$ each) (Figure 10a) has been determined. For better visualization the data was binned and a Gaussian fit was applied (Figure 10b). The resulting standard deviations in case of a living and a fixed cell are 21 nm and 9 nm, respectively. An analysis of a single time traces is shown in Figure 10a. Membrane fluctuation amplitudes turn out to amount to a few 10 nm and appear to be substantially larger on living than on fixed cells. Since PFA, via denaturation, only stiffens protein and leaves nucleic acid, lipids and aqueous medium intact, one may be tempted to conclude that ATP-driven or metabolic processes are responsible for the non-equilibrium character, i.e. larger extent of live cell fluctuations [52–54]. Especially active membrane motions generated via the spectrin corset underneath the membrane are suppressed if the cell is dead [55]. Membrane fluctuations of fixed cells are assumed to be merely thermally driven. Random samplings yield that osteoblast live-cell fluctuations exhibit about twice larger rms amplitudes than fixed. This is in line with the results on red blood cells [56], where membrane fluctuation amplitudes have been determined via holographic microscopy and found to be seven times larger on living versus fixed red blood cells in the frequency range of 0.2 to 12Hz. A rms amplitude 3.6 times higher on living than on fixed fibroblast-like cells (L929) was reported [37]. However, comparing the apparent fluctuations between living and fixed cells is not straight forward, since not only the bare membrane fluctuations are measured. It needs to be considered that macroscopic cell migration as well as the guided motion of the membrane protrusions itself lead to an increase of the measured fluctuation amplitudes. It has been shown that e.g. microvilli are moved up to 100 nm/s on the surface of A431 epidermal cells [57]. Taking this into account, the high amplitudes of several seconds periodicity could partly result from ruffles moving beneath the pipette opening. Slightly higher rms amplitudes may also be due to different temperature because fixed cells are measured at room temperature, while live cell imaging is done at 37 °C.

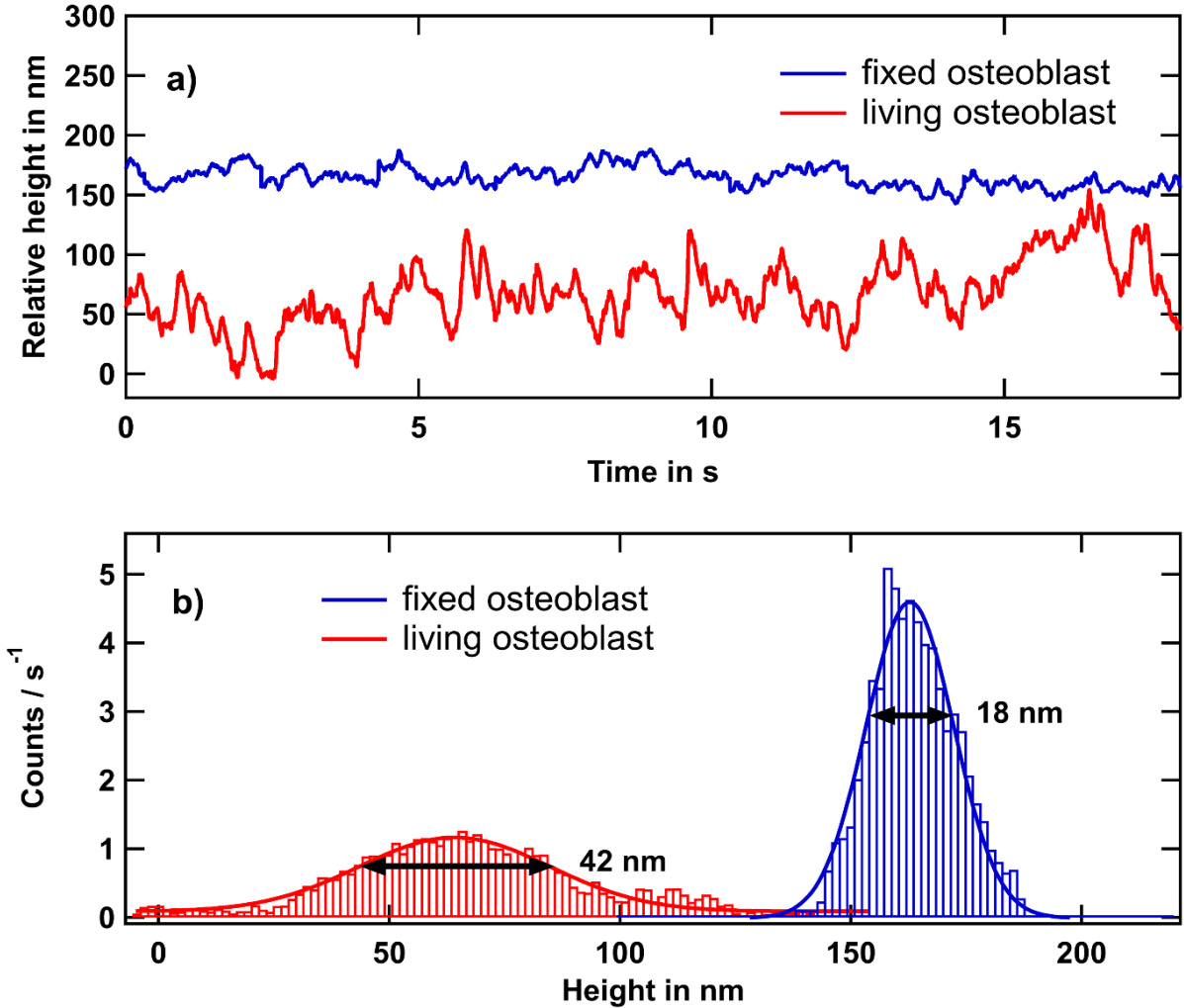

Figure 10: Membrane fluctuations of fixed and living MG-63 osteoblastic cells. a) Local height variations over time of membrane displacements of living (red) and fixed (blue) cells. b) Histograms and corresponding gauss fits of the traces yield standard deviations of 21 nm and 9 nm, respectively. The two distributions were shifted against each other horizontally for better visibility.

## Frequency response behavior

Owing to the fact that morphological changes at the cell surface and cell migration happen rather slowly, they should only contribute to low-frequency fluctuations. To investigate the frequency behavior we focus on measuring the height with activated feedback-loop in the frequency range of 0.2 to 500 Hz. Most systems exhibit a $f^{-m}$ spectral power density (SPD) with m varying between 4/3 to 2, depending on frequency band [58]. For living eukaryotic cells lower exponents (i.e. higher m) have been derived [53, 59, 60]. Indeed, for osteoblasts we observed $m = 2.48 \pm 0.18$ in the frequency regime of 0.1 to 1 Hz (see Figure 11 exemplarily). Beyond 1 Hz to ~10 Hz m becomes around 4/3 which was associated to the physical effect of hydrodynamic bending of the membrane [59]. As reference we used glass substrates and fixed osteoblasts, which exhibited m values of 2 in accordance with viscous stress of the liquid [58]. Subtle deviations from $f^{-m}$ sometimes can be observed, in terms of varying exponents at low frequencies for long time periods, beyond a couple of seconds. Those may originate from migration, however extensive statistics taking into the account location on the spread cell are required. Nevertheless, extracting the fraction of active, i.e. ATP-depending fluctuations, with spatial and frequency band resolution, may help to develop cell activity parameters for the assessment for cellular programs such as adhesion on material surfaces.

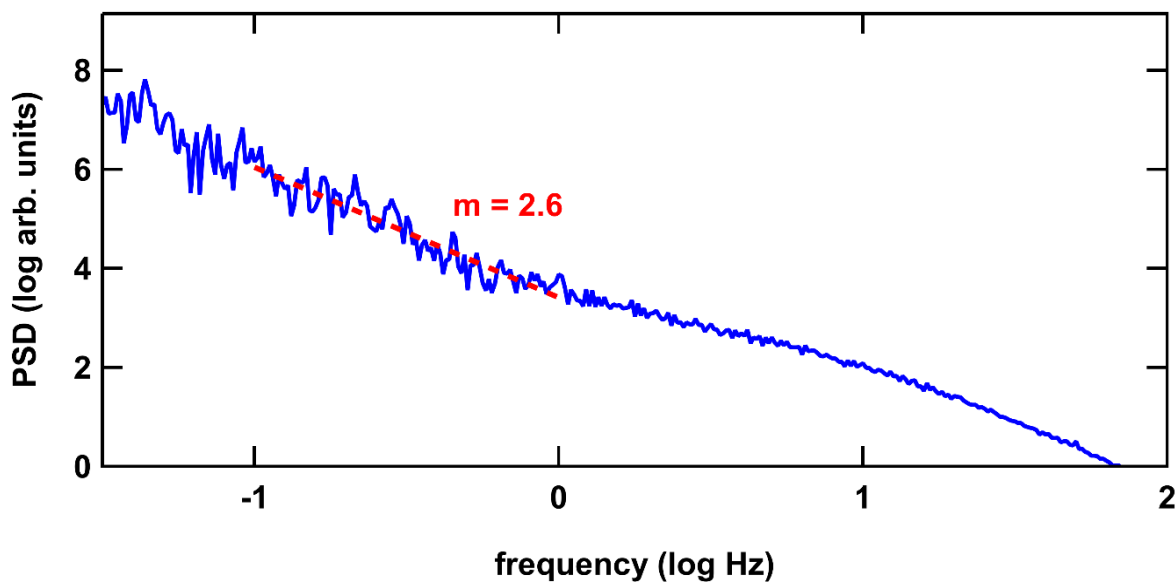

Figure 11: Power spectral density of vertical membrane fluctuations of a living osteoblastic cell. The exponential decay of the spectral density in the regime of 0.1 Hz to 1 Hz amounts to m = 2.6.

# Conclusion

High resolution SICM topographies of living and fixed MG-63 osteoblasts reveal three-dimensional morphology and dynamics of their cell membranes under largely noninvasive conditions. The most striking feature observed is the large amount of sheet like plasma membrane protrusions, i.e. ruffles of dorsal and peripheral type. Dorsal ruffles vanish to a large extend at cell rim regions when spreading is promoted and cells exhibit large adhesion areas (>2500 µm²), i.e. on amine-covered substrates, leading to smoother regions towards lamellipodia. This suggests that ruffles act as membrane storage. However, we cannot discriminate between the scenarios whether the cell anticipates rapid adhesion, i.e. ruffled membrane is brought towards the cell rim in order to establish lamellipodia or whether ruffles remain behind after unsuccessful lamellipodia formation. Yet the former scenario is rather unlikely since cells in the early stages of adhesion (low adhesion area) on PPAAm coating did not show higher ruffle density than those on glass. This means there is no sign of higher ruffle production or exposure prior to adhesion when the cell anticipates rapid spreading.

Heights at the osteoblastic cell's rim edges vary between 100 nm and more than 2 µm along the circumference. The smallest heights of around 60 nm may be due to ECM material left behind after the cell has retracted from such positions. Small heights of a few 100 nm may refer to lamellipodia regions while the larger heights of a few µm and less curvature are typical for ordinary configuration at cell rims.

Membrane fluctuation analyses reveal larger displacement amplitudes for living than for fixed cells. Looking at the frequency space of membrane fluctuations, living osteoblast show slightly higher absolute scaling exponents of the PSD compared to fixed ones. However, up to now it is not clear whether these characteristics are specific for different locations e.g. the cell nucleus or lamellipodia or if they even differ for various substrates offered to the cells which would mean these exponents are cell program specific.

# Experimental

## Substrate preparation and characterization

Either pristine borosilicate glass or coatings of positively charged plasma-polymerized allylamine (PPAAm) or negatively charged Au was used, see below [61]. Zeta potential and water contact angle were measured to be 8.6 mV and 68° for PPAAm, respectively [62]. In case of Au-layer -119 mV and 101° were determined, respectively.

ζ-potential measurements were performed using the SurPASS™ system (Anton Paar, Ostfildern, Germany) to determine the surface potential. Au- and PPAAm modified titanium substrates were placed in pairs in the measuring chamber with a gap height of 100 µm. The streaming potential was measured at pH 6.5 to 8.0, at 150 mbar in a 1 mM KCl solution (VWR International, Darmstadt, Germany). ζ potentials were determined with the associated software Attract 2.1 (Anton Paar, Ostfildern, Germany) according to the Helmholtz-Smoluchowski equation. ζ potentials at pH 7.4 were calculated with the function 'linear regression' using the software GraphPad Prism Version 6.05 (n = 3).

Water contact angle (WCA) values were obtained by the sessile drop method using the Drop Shape Analyzer—DSA25 (Krüss, Hamburg, Germany). Drop shape images of 1 µl water drops were acquired with the digital camera of the DSA25 under atmospheric condition at room temperature (n = 3). WCA values were calculated with the associated software (ADVANCE, V.1.7.2.1, Krüss, Hamburg, Germany) via Young´s equation.

Plasma polymer coating: The specimens were coated with a plasma polymerized allylamine (PPAAm) nanolayer by using a low-pressure plasma reactor (V55G, Plasma Finish, Germany) according to the two-step procedure described earlier [1, 9]. Briefly, PPAAm was deposited by a microwave (MW)-excited (2.45 GHz), pulsed plasma (500 W, 50 Pa, 50 sccm Ar) for 480 s (effective treatment time to initiate plasma polymerization of allylamine).

For Au coating a sputter coater (Cressington 108 auto/SE, UK) was used with the following parameters: Ar pressure: 0.06 mbar, sputter current: 30 mA, time: 60 s, distance sample-target: 55 mm. The thickness of the coating was measured online by a quartz crystal thickness monitor (Cressington MTM 10, UK) and the sputter process was stopped at a nominal value of approximately 10 nm. The Au layers have only been applied to half of the glass area in order to create an in-situ reference. Since we observed spread cells adhering directly over the border between Au and glass the cells apparently have no preference.

## Cell culture

For the SICM experiments, human osteoblast-like cells of the cell line MG-63 (American Type Culture Collection ATCC®, CRL1427™, Bethesda, USA) were used. This cell line has been successfully applied for studying cell-material interactions [63] with similar characteristics to primary human osteoblasts [64, 65]. Cells were cultured in Dulbecco's modified Eagle's medium (DMEM, 31966-021, Life Technologies Limited, Paisley, UK), with 10% fetal calf serum (FCS, Biochrom FCS Superior, Merck, Darmstadt, Germany) and 1% antibiotics (gentamicin, Ratiopharm, Ulm, Germany) in a humidified atmosphere at 37 °C with 5% $CO_2$. The differently covered borosilicate glass slides (22mm x 22mm) were cleaned with ethanol. Cells were seeded with a density of 4000/cm² and cultured for 24 h. This rather low density largely guarantees single cells which are desirable for investigation of cell edge heights.

For comparison and resolution enhancement in SICM analysis the cells of some samples were fixed. In that case 4% PFA (Sigma-Aldrich, St. Louis, MO, USA) has been applied for 10 minutes at room temperature.

## Measurement principle and data preparation

Scanning electron microscopy (SEM) has been performed using a field emission SEM (Gemini Supra 25, Zeiss) at 1 keV electron energy without Au coating.

For SICM a commercial AFM/SICM setup (NX-bio, Park Systems, Korea) with a live-cell chamber (5% $CO_2$, 37°C) has been used. The sample was immersed in physiological electrolyte (DMEM with 10% FCS and 1% gentamicin) in case of living topography and membrane fluctuation measurements. Measurements of fixed cells took place at room temperature (21°C) in phosphate-buffered saline (PBS).

Nanopipettes with opening diameters below 80 nm have been pulled from borosilicate capillary tubes (inner diameter 0.58 mm) using a $CO_2$ laser puller (Sutter P-2000, USA). The following parameters have been used: Heat: 260, Filament: 4, Velocity: 50, Delay: 225, Pull: 140. The opening diameter has been exemplarily determined by SEM and via I-V characteristics [19].The bias voltage applied between pipette and bath electrode was in the regime of 100 mV. Both electrodes are non-polarizable (Ag/AgCl).

For SICM topography measurements the nanopipette is approaching the surface until the setpoint of 0.98 nA (corresponding to ion current reduction of 2%) is reached. After that the pipette is retracted for a couple of µm with respect to the latest acquired surface height and moved laterally to the next scanning point. The reference current is measured, followed by re-approaching until the setpoint is reached. This is referred to as hopping, approach-retract or dynamic scanning to avoid crashes at steep edges due to sideway insensitivity of the pipette. The error signal, that is the difference between setpoint and actual ion current, should be small and is taken as co-image. For membrane dynamics measurements the nanopipette is kept at constant lateral position over the cell and either temporal height variations with activated feedback loop or current variations at deactivated feedback loop are acquired. Temporal current and height spectra have been evaluated by Igor Pro (WaveMetrics, Inc)

As main observables we used the nanomorphology and dynamics of lamellipodia and rim features of adhered osteoblasts in the initial phase of cellular adhesion. SICM data evaluation has been performed using Gwyddion and Igor Pro. A plane has been subtracted from the raw data to cancel out sample tilt and long-term drift. The different rows along the fast scan direction have been aligned by adjusting the row offset such that the median value of the difference between two neighboring rows equals zero. Images shown in Figure 2a and 2e have been further processed by subtracting 2D-polynomial backgrounds using Igor Pro to emphasize small features. Furthermore, the glass surface shown in Figure 2a as a black area on the right side was artificially set to height "0". Colors in 3D rendered images are a combination of simulated illumination and height information by using Gwyddion's 3D view (option overlay checked). Surface areas and projected areas have been determined using Gwyddion's built-in statistical analysis feature.

Frequency behavior plots have been evaluated from raw data using the FFT-algorithm with a Blackman window in Igor Pro. Optical microscopy surveys have been acquired by an inverted microscope (Nikon ECLIPSE Ti-U, Japan) instrument from below through the glass slide. Since the lateral SICM frame sizes are rather small an optical overview image was taken from below, in order to place the nanoprobe at the selected position from top. After that, light is switched off to keep potential field triggers low. A second optical image was taken after SICM acquisition, in order to ensure spatial assignment in case of migrating cells.

# Acknowledgements

This work was supported by the German Research Foundation (Deutsche Forschungsgemeinschaft, DFG) within the Collaborative Research Centre 1270 ELAINE as well as the European Regional Development Fund (EFRE). We thank Björn Vogler, Henrike Rebl, Moritz Hofheinz, Thomas Freitag for their help with analyses, pipette pulling and cell culturing and preparation.